\documentclass[aps,reprint,superscriptaddress,nofootinbib]{revtex4-1}

\usepackage{bm, amsfonts, amsmath, amssymb, xspace, mathtools, amsthm}
\usepackage{graphicx}
\usepackage{hyperref}
\usepackage[dvipsnames]{xcolor}
\newcommand\myshade{85}
\colorlet{mylinkcolor}{BrickRed}
\colorlet{mycitecolor}{NavyBlue}
\colorlet{myurlcolor}{Aquamarine}
\hypersetup{
  linkcolor  = mylinkcolor!\myshade!black,
  citecolor  = mycitecolor!\myshade!black,
  urlcolor   = myurlcolor!\myshade!black,
  colorlinks = true,
}

\def\compiletikz{0}

\if1\compiletikz
\input{tikzsettings.tex}
\fi

\newcommand{\splitatcommas}[1]{%
\begingroup
\begingroup\lccode`~=`, \lowercase{\endgroup
    \edef~{\mathchar\the\mathcode`, \penalty0 \noexpand\hspace{0pt plus 1em}}%
    }\mathcode`,="8000 #1%
    \endgroup
}

\begin{document}

\title{Disentangling high-order mechanisms and high-order behaviours in complex systems}

\author{Fernando E. Rosas}
\thanks{F.R., P.M. and A.L. contributed equally to this work.\\Correspondence: \texttt{f.rosas@ic.ac.uk}, \texttt{\{pam83,al857\}@cam.ac.uk}}
\affiliation{Data Science Institute, Imperial College London, UK}
\affiliation{Centre for Psychedelic Research, Imperial College London, UK}
\affiliation{Centre for Complexity Science, Imperial College London, UK}

\author{Pedro A.M. Mediano}
\thanks{F.R., P.M. and A.L. contributed equally to this work.\\Correspondence: \texttt{f.rosas@ic.ac.uk}, \texttt{\{pam83,al857\}@cam.ac.uk}}
\affiliation{Department of Psychology, University of Cambridge, UK}
\affiliation{Department of Psychology, Queen Mary University of London, UK}

\author{Andrea I Luppi}
\thanks{F.R., P.M. and A.L. contributed equally to this work.\\Correspondence: \texttt{f.rosas@ic.ac.uk}, \texttt{\{pam83,al857\}@cam.ac.uk}}
\affiliation{University Division of Anaesthesia, University of Cambridge, UK}
\affiliation{Department of Clinical Neurosciences, University of Cambridge, UK}
\affiliation{Leverhulme Centre for the Future of Intelligence, University of Cambridge, UK}

\author{Thomas F. Varley}
\affiliation{Department of Psychological \& Brain Sciences, Indiana University, Bloomington, IN, USA}
\affiliation{School of Informatics, Computing, and Engineering, Indiana University, Bloomington, IN, USA}

\author{Joseph~T.~Lizier}
\affiliation{Centre for Complex Systems and Faculty of Engineering, The University of Sydney, Australia}

\author{Sebastiano Stramaglia}
\affiliation{Dipartimento Interateneo di Fisica, Università degli Studi di Bari Aldo Moro, Italy}
\affiliation{INFN, Sezione di Bari, Italy}

\author{Henrik J. Jensen}
\affiliation{Centre for Complexity Science, Imperial College London, UK}
\affiliation{Department of Mathematics, Imperial College London, UK}
\affiliation{Institute of Innovative Research, Tokyo Institute of Technology, Japan}

\author{Daniele Marinazzo}
\affiliation{Department of Data Analysis, Ghent University, Belgium}

\maketitle


Battiston \textit{et al.}~\cite{battiston2021physics} provide a comprehensive overview of how investigations of complex systems should take into account interactions of more than two elements, which can be modelled by hypergraphs and studied via topological data analysis. Following a separate line of enquiry, a broad literature has developed information-theoretic tools to characterise high-order interdependencies from observed data. While these could seem to be competing approaches aiming to address the same question, in this correspondence we clarify that this is \emph{not} the case, and that a complete account of higher-order phenomena needs to embrace both.

\begin{figure*}[ht]
    \centering
    \includegraphics[width=0.8\linewidth]{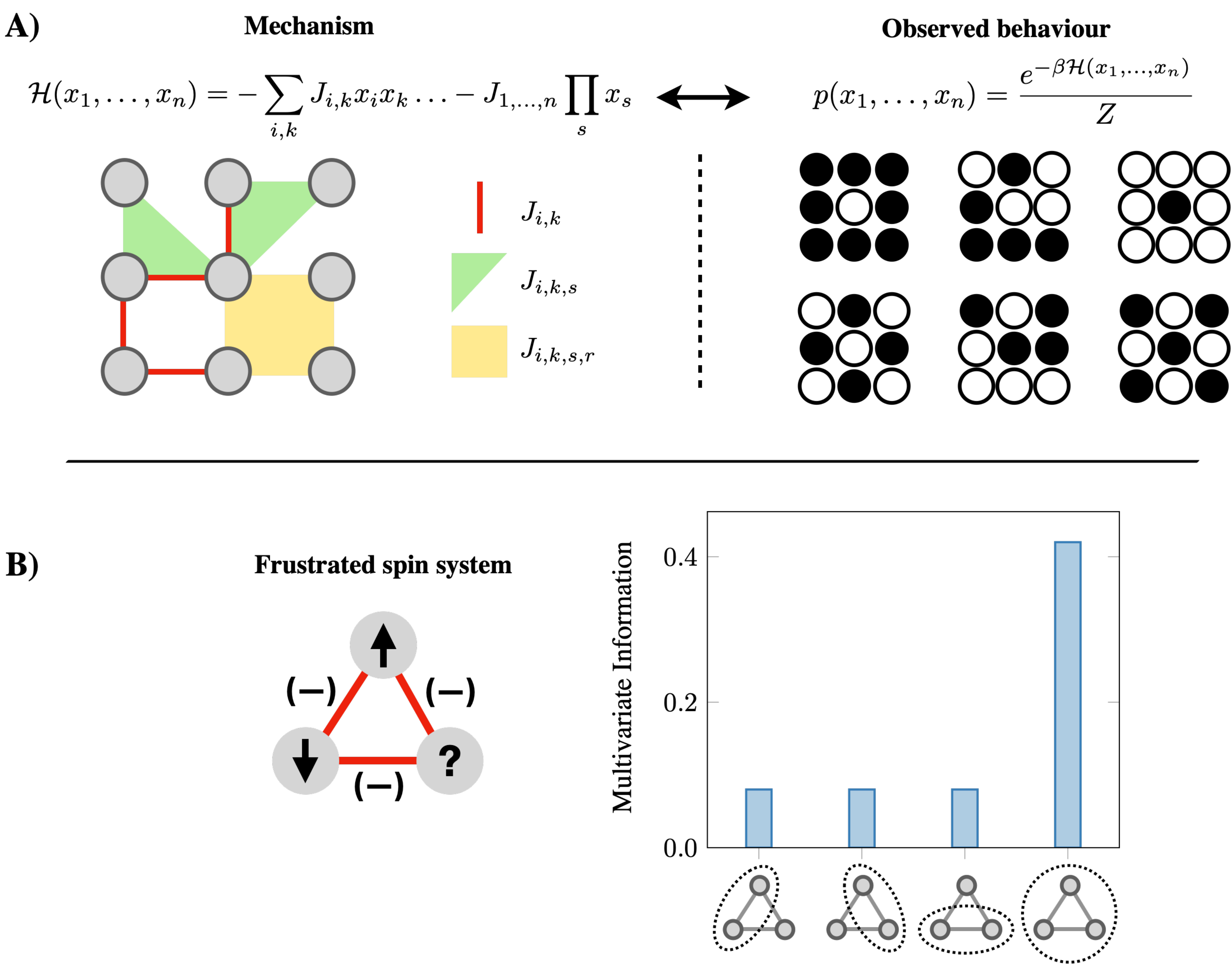}
    \caption{
    \textbf{A)} For a spin glass model, the mechanism of the system is given by a Hamiltonian that encodes (low-order) pairwise connection and (high-order) multiple interactions, forming a hypergraph. In contrast, the behaviour of the system is given by the frequency of the observed patterns encoded in the joint probability distribution -- which in turn can establish high-order interdependencies such as synergy or redundancy~\cite{williams2010nonnegative,lizier2018information,mediano2021towards,rosas2019quantifying}.
    \textbf{B)} A small frustrated spin model, in which the topology of the negative couplings makes it impossible to simultaneously satisfy the tendency of all spins to be different from their neighbours~\cite{matsuda2000physical}. A direct calculation shows that the total interdependency in the system (as measured by the \textit{Total Correlation}~\cite{rosas2019quantifying}) is significantly larger than the sum of the three pairwise interdependencies. This is an example of how low-order (pairwise) mechanisms can give rise to high-order behaviour that cannot be explained in terms of pairwise statistics.
}
    \label{fig:model}
\end{figure*}

The approaches reviewed by Battiston and colleagues put a special focus\footnote{We highlight this as a dominant trend through the paper, while acknowledging that some of the considered techniques can be used to represent both mechanisms and behaviours.} on what could be described as \textit{high-order mechanisms}, which are usually modelled via Hamiltonians or dynamical laws involving beyond-pairwise interactions. A distinct, but complementary perspective is to focus on the resulting \textit{high-order behaviour}, i.e. patterns of activity that can be explained in terms of the whole but not the parts. Put simply, high-order mechanisms refer to the modelling of the data-generating process, while high-order behaviours refer to emergent properties within the resulting multivariate statistics. For example, in a spin glass system, high-order causes would correspond to high-order terms in the Hamiltonian, while high-order behaviours would refer to the emergent patterns resulting from the Boltzmann distribution (Figure~\ref{fig:model}A).

The investigation of high-order behaviour in data has a long history, stemming from foundational work in information theory~\cite{mcgill1954multivariate,watanabe1960information} and its early applications in biophysics~\cite{gat1999synergy,tononi1994measure}. These early efforts led to the formalisation of high-order behaviour into the framework of partial information decomposition (PID) and its subsequent developments~\cite{williams2010nonnegative,lizier2018information,mediano2021towards}, which have established formal bases for the analysis of high-order interdependencies exhibited by groups of three or more variables, and a description of their (synergistic or redundant) nature. 

Mechanisms and behaviours address fundamentally different questions: the former address how the system is structured, while the latter focus on emergent properties related to what it “does.” Crucially, these questions are not interchangeable; intuition might suggest that high-order behaviour necessarily rests upon high-order mechanisms, but this is not the case (Figure~\ref{fig:model}B). Consequently, neglecting high-order behaviour risks missing important aspects of complex phenomena. Furthermore, when the goal is to identify high-order behaviour, high-order methods relying only on pairwise statistics (e.g. simplicial complexes built from a correlation matrix) may be in principle insufficient, as significant information can be present only in the joint probability distribution and not the pairwise marginals (Figure~\ref{fig:model}B).

In summary, differentiating between high-order mechanisms and behaviours allows greater precision in articulating hypotheses, and in choosing the right tools to probe them. This distinction will facilitate the collaboration between different communities devoted to the study of complex systems, as the complementarity of these perspectives will provide a more powerful and encompassing avenue for deepening our understanding of high-order phenomena.

\bibliographystyle{naturemag}
\bibliography{main}

\begin{thebibliography}{10}
\expandafter\ifx\csname url\endcsname\relax
  \def\url#1{\texttt{#1}}\fi
\expandafter\ifx\csname urlprefix\endcsname\relax\def\urlprefix{URL }\fi
\providecommand{\bibinfo}[2]{#2}
\providecommand{\eprint}[2][]{\url{#2}}

\bibitem{battiston2021physics}
\bibinfo{author}{Battiston, F.} \emph{et~al.}
\newblock \bibinfo{title}{The physics of higher-order interactions in complex
  systems}.
\newblock \emph{\bibinfo{journal}{Nature Physics}}
  \textbf{\bibinfo{volume}{17}}, \bibinfo{pages}{1093--1098}
  (\bibinfo{year}{2021}).

\bibitem{mcgill1954multivariate}
\bibinfo{author}{McGill, W.}
\newblock \bibinfo{title}{Multivariate information transmission}.
\newblock \emph{\bibinfo{journal}{Transactions of the IRE Professional Group on
  Information Theory}} \textbf{\bibinfo{volume}{4}}, \bibinfo{pages}{93--111}
  (\bibinfo{year}{1954}).

\bibitem{watanabe1960information}
\bibinfo{author}{Watanabe, S.}
\newblock \bibinfo{title}{Information theoretical analysis of multivariate
  correlation}.
\newblock \emph{\bibinfo{journal}{IBM Journal of research and development}}
  \textbf{\bibinfo{volume}{4}}, \bibinfo{pages}{66--82} (\bibinfo{year}{1960}).

\bibitem{gat1999synergy}
\bibinfo{author}{Gat, I.} \& \bibinfo{author}{Tishby, N.}
\newblock \bibinfo{title}{Synergy and redundancy among brain cells of behaving
  monkeys}.
\newblock \emph{\bibinfo{journal}{Advances in neural information processing
  systems}} \bibinfo{pages}{111--117} (\bibinfo{year}{1999}).

\bibitem{tononi1994measure}
\bibinfo{author}{Tononi, G.}, \bibinfo{author}{Sporns, O.} \&
  \bibinfo{author}{Edelman, G.~M.}
\newblock \bibinfo{title}{A measure for brain complexity: relating functional
  segregation and integration in the nervous system}.
\newblock \emph{\bibinfo{journal}{Proceedings of the National Academy of
  Sciences}} \textbf{\bibinfo{volume}{91}}, \bibinfo{pages}{5033--5037}
  (\bibinfo{year}{1994}).

\bibitem{williams2010nonnegative}
\bibinfo{author}{Williams, P.~L.} \& \bibinfo{author}{Beer, R.~D.}
\newblock \bibinfo{title}{Nonnegative decomposition of multivariate
  information}.
\newblock \emph{\bibinfo{journal}{arXiv preprint arXiv:1004.2515}}
  (\bibinfo{year}{2010}).

\bibitem{lizier2018information}
\bibinfo{author}{Lizier, J.~T.}, \bibinfo{author}{Bertschinger, N.},
  \bibinfo{author}{Jost, J.} \& \bibinfo{author}{Wibral, M.}
\newblock \bibinfo{title}{Information decomposition of target effects from
  multi-source interactions: Perspectives on previous, current and future work}
  (\bibinfo{year}{2018}).

\bibitem{mediano2021towards}
\bibinfo{author}{Mediano, P.~A.} \emph{et~al.}
\newblock \bibinfo{title}{Towards an extended taxonomy of information dynamics
  via integrated information decomposition}.
\newblock \emph{\bibinfo{journal}{arXiv preprint arXiv:2109.13186}}
  (\bibinfo{year}{2021}).

\bibitem{rosas2019quantifying}
\bibinfo{author}{Rosas, F.~E.}, \bibinfo{author}{Mediano, P.~A.},
  \bibinfo{author}{Gastpar, M.} \& \bibinfo{author}{Jensen, H.~J.}
\newblock \bibinfo{title}{Quantifying high-order interdependencies via
  multivariate extensions of the mutual information}.
\newblock \emph{\bibinfo{journal}{Physical Review E}}
  \textbf{\bibinfo{volume}{100}}, \bibinfo{pages}{032305}
  (\bibinfo{year}{2019}).

\bibitem{matsuda2000physical}
\bibinfo{author}{Matsuda, H.}
\newblock \bibinfo{title}{Physical nature of higher-order mutual information:
  Intrinsic correlations and frustration}.
\newblock \emph{\bibinfo{journal}{Physical review E}}
  \textbf{\bibinfo{volume}{62}}, \bibinfo{pages}{3096} (\bibinfo{year}{2000}).

\end{thebibliography}

\end{document}